\documentstyle[aps,prb,multicol,epsf]{revtex}
\newcommand{\lineright}{\rule[-1ex]{0.1mm}{1ex}\rule{0.485\linewidth}{0.1mm}}
\newcommand{\lineleft}{\rule{0.485\linewidth}{0.1mm}\rule[0mm]{0.1mm}{1ex}}
\begin{document}
\draft
\title{Andreev Bound States at the Interface of
  Antiferromagnets and d-wave Superconductors}
\author{Brian M\o ller Andersen and Per Hedeg\aa rd}
\address{\O rsted Laboratory, Niels Bohr Institute for APG,
Universitetsparken 5, DK-2100 Copenhagen \O\, Denmark}
\date{June 17 2002}
\maketitle
\begin{abstract}
We set up a simple transfer matrix formalism to study the existence of
bound states at interfaces and in junctions between antiferromagnets and
d-wave superconductors. The well-studied zero energy mode at the
\{110\} interface between an insulator and a d$_{x^2-y^2}$ wave
superconductor is spin split when the insulator is an antiferromagnet.
This has as a consequence that any competing interface induced superconducting
order parameter that breaks the time reversal symmetry needs to exceed a
critical value before a charge current is induced along the
interface.
\end{abstract}
\pacs{PACS numbers: 74.50+r, 74.72.-h, 74.25.Ha}
\begin{multicols}{2}
The discovery of the symmetry of the superconducting order parameter has
been one of the most successful studies of the High-T$_c$
materials. Angular resolved photoemission spectroscopy has revealed the
nodes in the gap function and tunneling experiments have proven the
sign change between adjacent lobes of the d$_{x^2-y^2}$ wave
gap\cite{shen,harlingen,kashiwaya}. It
was first shown by Hu\cite{hu} that this sign change can lead to
zero energy Andreev bound states (ZEBS) at the surface
of an insulator and a d-wave superconductor. These Andreev bound states were
later identified with the zero bias conductance peaks observed in
tunneling experiments. The experiments by Covington
{\sl et.al.}\cite{covington} indicated, however, that the
surface states were spontaneously split by a minigap. Several ideas
were proposed for this effect\cite{lofwander}; one of which included
the induction of a time reversal symmetry breaking $is$ component of
the order parameter near the interface\cite{fogelstrom}. The resulting
gap $d+is$ lowers the condensation energy by 
lifting the directional degeneracy of the ZEBS\cite{sigrist}. Later
Honerkamp {\sl et.al.}\cite{honerkamp} used a tight-binding model
with onsite repulsion and spin dependent nearest neighbor interaction
to self-consistently study the competition between additional induced
orders near the surface of an insulator and a d$_{x^2-y^2}$ wave
superconductor.\\
Motivation for studying close domains of antiferromagnetism and
superconductivity arises from the existence of striped domains in the
cuprate materials. This was further emphasized by recent elastic neutron scattering
experiments showing that static antiferromagnetic order is induced in a superstructure around
the vortices in the mixed state of La$_{2-x}$Sr$_x$CuO$_4$\cite{bella} and
La$_2$CuO$_{4+\delta}$\cite{khaykovich}. These experiments are
consistent with a static environment of alternating antiferromagnetic
and d-wave superconducting stripes around the vortex cores. Thus the
electronic states in such an environment is an important question.\\ 
Inspired by these experiments we set up a
simple transfer matrix method to identify bound states on interfaces
and junctions between antiferromagnets and d-wave superconductors. In
particular we discuss a single interface separating
antiferromagnetic and d-wave superconducting half-planes (AF/dSC), and
point out a few differences from the conventional non-magnetic insulator-d-wave
superconductor interface (I/dSC). Note that the antiferromagnetism
forces us to study a lattice model which is contrary to the usual
discussion of Andreev interference in terms of semi-classical
continuum models.\\  
A simple lattice model that includes both d-wave superconductivity and
antiferromagnetism is given by the following Hamiltonian
\begin{eqnarray}
H &=&-t \sum_{\left< n,m \right>\sigma}c^\dagger_{n\sigma} c_{m\sigma}
+ H.c. -\mu \sum_{n\sigma}
c^\dagger_{n\sigma} c_{n\sigma} \\
&+& \sum_{\left< n,m \right>} \Delta_{n,m}
c^\dagger_{n\uparrow} c^\dagger_{m\downarrow} +H.c. \\
&+&\sum_{n} M_n
\left(c^\dagger_{n\uparrow} c_{n\uparrow} - c^\dagger_{n\downarrow} c_{n\downarrow} \right)
\end{eqnarray}
where $\left< n,m \right>$ denotes nearest neighbors. $M_n$ and
$\Delta_{n,m}$ are the spatially dependent magnetic and
superconducting order parameters. This Hamiltonian is quadratic and can
be diagonalized by a Bogoliubov-de Gennes (BdG) transformation
\begin{equation}
\gamma^\dagger_\sigma = \sum_n u_{\sigma}(n)
c^\dagger_{n\sigma} + \sigma v_{\sigma} (n) c_{n-\sigma}.
\end{equation}
with $\sigma$ equal to +1 (-1) for spin up (down).
We use the notational convention that the spin indices on $u_\sigma$
and $v_\sigma$ follow that on the Bogoliubov operators
$\gamma^\dagger_\sigma$.\\
In the case of a $d_{x^2-y^2}$-wave superconductor there are two
qualitatively different orientations of
the interface; the \{100\} and \{110\} directions corresponding to a
vertical and diagonal stripe respectively. Both cases are
studied below with the $x$-axis ($y$-axis) chosen 
perpendicular (parallel) to the interface which is placed at
$x=0$. The lattice constant is set to unity.
Assuming translational invariance along the $y$-direction the AF/dSC interface
reduces to a one dimensional problem. For the \{100\} interface the
resulting Bogoliubov-de Gennes equations have the form
\end{multicols}
\noindent \lineleft
\begin{eqnarray}\label{bdg1a}
\epsilon_\sigma u_{q\sigma}\left(x\right)&=&-t \left(
  u_{q\sigma}\left(x+1\right) + u_{q\sigma}\left(x-1\right) + 2
  \cos(q) u_{q\sigma}\left(x\right) \right) - \mu
  u_{q\sigma}\left(x\right) + \sigma M_x
  u_{q+Q\sigma}\left(x\right) \\ \nonumber
&+& \left( \Delta^d_{x+1,x} \right)
  v_{q\sigma}\left(x+1\right) + \left( \Delta^d_{x-1,x} \right)
  v_{q\sigma}\left(x-1\right) + 2 \cos(q) \left(- \Delta^d_x \right)
  v_{q\sigma}\left(x\right) \\ \label{bdg1b}
\epsilon_\sigma v_{q\sigma}\left(x\right)&=&t \left(
  v_{q\sigma}\left(x+1\right) + v_{q\sigma}\left(x-1\right) + 2
  \cos(q) v_{q\sigma}\left(x\right) \right) + \mu
  v_{q\sigma}\left(x\right) + \sigma M_x
  v_{q+Q\sigma}\left(x\right) \\ \nonumber
&+& \left( \Delta^{*d}_{x+1,x} \right)
  u_{q\sigma}\left(x+1\right) + \left( \Delta^{*d}_{x-1,x} \right) 
  u_{q\sigma}\left(x-1\right) + 2 \cos(q) \left(- \Delta^{*d}_x \right)
  u_{q\sigma}\left(x\right)
\end{eqnarray}
\hfill \lineright
\begin{multicols}{2}
\noindent after fourier transforming along the $y$ direction. The corresponding
equations for the fourier components $u_{q+Q\sigma}$ and
$v_{q+Q\sigma}$ are obtained by 
simply performing the substitution $q \rightarrow q+Q$. These BdG
equations are diagonal in the spin index with the only difference
between spin up and down being the sign of the magnetic term.\\
A simple way to study bound states at the interface is in
terms of the transfer matrix method\cite{merzbacher}. Thus we introduce a
($q,\epsilon$)-dependent matrix $T\left(x+1,x\right)$ defined by
\begin{equation}
\Psi\left(x+1\right) = T\left(x+1,x\right) \Psi\left(x\right).
\end{equation}
which transfers the spinor $\Psi$ from site $x$ to site $x+1$. For a
model with nearest neighbor coupling $\Psi$ takes the explicit form
$\Psi \left(x\right) = \left( \psi \left(x\right),\psi \left(x-1\right) \right)$ where 
\begin{equation}
\psi \left(x\right) = \left( u_{q\sigma}\left(x\right),
  v_{q\sigma}\left(x\right), u_{q+Q\sigma}\left(x\right),
  v_{q+Q\sigma}\left(x\right) \right).
\end{equation} 
The associated $8\times8$ transfer matrix has the general form
\begin{equation}
T\left(x+1,x\right) = 
\left(
\begin{array}{cc}
A & B \\
1 & 0
\end{array}
\right)
\end{equation} 
where $A$ ($B$) denotes the $4\times4$ coefficient-matrix connecting
$\psi\left(x+1\right)$ and $\psi\left(x\right)$ ($\psi\left(x-1\right)$)
determined from the BdG equations (\ref{bdg1a}-\ref{bdg1b}). In the
simplest case of a sharp interface 
we have the following spatial dependence of $M_x$ and
$\Delta_x$
\begin{eqnarray}
M_x &=& M \left(-1\right)^x~~~~~~\mbox{for}~~~~x \leq 0 \\
\Delta_x &=& \Delta_d~~~~~~~~~~~~~~\mbox{for}~~~~x > 0 
\end{eqnarray}
Thus there are effectively three different transfer matrices; one in the bulk
magnetic region $T_M$, one in the bulk superconducting region $T_{SC}$
and one associated with transfer through the interface $T_I$. By
diagonalizing $T_M$ and $T_{SC}$ there exists decaying, growing or
propagating eigenstates depending on whether the eigenvalues are less,
larger or equal to one, respectively. Here, decaying and growing are
referred to propagation along the $x$-axis for increasing $x$. If
$PET_M$ denotes the matrix obtained after propagating the eigenvectors
of the bulk magnetic transfer matrix through the interface we
introduce a matrix $\alpha$ given by 
\begin{equation}
PET_M = ET_{SC}.\alpha
\end{equation} 
where $ET_{SC}$ is the matrix containing the eigenvectors
of the bulk superconducting region as coloum vectors. The dot
indicates matrix multiplication. Now, let $S_g^{m}$ and $S_g^{sc}$
denote the subspace of growing eigenstates of $PET_M$ and $ET_{SC}$
respectively,
and consider the following linear combination of the {\sl growing}
states of $PET_M$ 
\begin{eqnarray}\label{olebole}
\sum_{i\in S_g^{m}} \beta_i |PET_{M}i> &=& \sum_{i\in S_g^{m}} \sum_{j
  \in S_g^{sc}} \beta_i \alpha_{ji} |ET_{SC}j> \\ \nonumber 
&=&  \sum_{j\in S_g^{sc}} \left( \sum_{i\in
  S_g^{m}} \alpha_{ji} \beta_i \right) |ET_{SC}j> 
\end{eqnarray}
From equation (\ref{olebole}) it is evident that to have a bound state at
the interface the vector $\beta$ must belong to the null space of
the reduced matrix $\alpha_r$, which is the $S_g^{sc} \times S_g^{m}$ upper
left part of the original matrix $\alpha$ since the
matrices $PET_M$ and $ET_{SC}$ are organized to have the eigenstates with the
largest eigenvalues as coloum vectors to the left. In the case that
the two subspaces $S_g^{sc}$ and 
$S_g^m$ have the same dimension a bound state at the interface is
characterized by the vanishing of the determinant of $\alpha_r$
\begin{equation}\label{det}
\mbox{Bound states:}~~~~~~~\det\left(\alpha_r\right) = 0.
\end{equation}
Plots of the wavefunctions with values of ($q,\epsilon$) that satisfy
Eqn. (\ref{det}) verifies that these states indeed are bound to the
interface (not shown).
The following explicit values of the input parameters are chosen: $t=1$,
$\Delta_d=0.14$, $M=2.0$ and $\mu=-0.99$ (for simplicity we ignore
next-nearest neighbor coupling).  
Figure 1a shows the determinant plottet as a function of energy for
the \{100\} interface. There 
are bound states close to the superconducting gap edge that disperses
downward in a cosine form (Figure 1b). These are the well-known de
Gennes/Saint-James states existing on the surface of an insulator and a
superconductor\cite{deGennes,comment}. 
\begin{figure}
\centerline{\epsfxsize=\linewidth\epsfbox{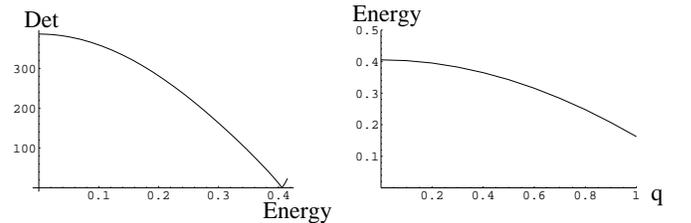}}
\caption{\{100\} interface between an antiferromagnet and a d-wave
  superconductor: a) Determinant of $\alpha_r$ as a function of energy
  $\epsilon$ for $q=0.1$. There is a de Gennes/Saint-James bound state
  close to the superconducting gap edge. As seen in b), their dispersion
  has the expected downward cosine form until it merges with the continuum.}   
\end{figure}
The induction of additional gap
symmetries, extended s-wave or p-wave, near the \{100\} interface of a
d-wave superconductor and an antiferromagnet has
been studied self-consistently by Kuboki\cite{kuboki}. These local gap
perturbations will slightly modify the graphs in Figure 1. There is no
spin splitting of the dGSJ mode in this geometry.\\ 
We turn now to the more interesting configuration of a \{110\}
interface. Allowing for a possible interface induced sub-gap order
with extended s-wave symmetry the Bogoliubov-de Gennes equations have the
form 
\end{multicols}
\noindent \lineleft
\begin{eqnarray}
\epsilon_\sigma u_{q\sigma}\left(x\right)&=& -2t \cos(p) \left(
  u_{q\sigma}\left(x+1\right) + u_{q\sigma}\left(x-1\right) \right) - \mu
  u_{q\sigma}\left(x\right) + \sigma M_x
  u_{q\sigma}\left(x\right) \\ \nonumber
&-& 2 i \sin(q) \left( \Delta^d_{x+1,x}
  v_{q\sigma}\left(x+1\right) - \Delta^d_x v_{q\sigma}\left(x-1\right)
  \right) + 2 i \cos(q) \left( \Delta^s_{x+1,x} v_{q\sigma}\left(x+1\right) + \Delta^s_{x}
  v_{q\sigma}\left(x-1\right) \right) \\
\epsilon_\sigma v_{q\sigma}\left(x\right)&=&2t \cos(p) \left(
  v_{q\sigma}\left(x+1\right) + v_{q\sigma}\left(x-1\right) \right) + \mu
  v_{q\sigma}\left(x\right) + \sigma M_x
  v_{q\sigma}\left(x\right) \\ \nonumber
&-& 2 i \sin(q) \left( \Delta^{*d}_{x+1,x}
  u_{q\sigma}\left(x+1\right) - \Delta^{*d}_x
  u_{q\sigma}\left(x-1\right) \right) - 2 i \cos(q) \left(
  \Delta^{*s}_{x+1,x} u_{q\sigma}\left(x+1\right) + \Delta^{*s}_x
  u_{q\sigma}\left(x-1\right) \right)
\end{eqnarray}
\hfill \lineright
\begin{multicols}{2}
\noindent These equations are diagonal in the fourier component $q$ obtained
after fourier transforming parallel to the \{110\} interface since
there is no staggering of the moments along a diagonal line in a square
antiferromagnetic lattice. In Figure 2 we plot again the determinant
of the reduced matrix 
$\alpha_r$ as a function of energy $\epsilon$ when $\Delta^s=0$. As seen
the spin degeneracy of the ZEBS (dashed curve) is lifted at a \{110\}
AF/dSC interface. As
opposed to the usual dGSJ states in Figure 1, this splitting 
is also caused by the fact that a \{110\} interface belongs to
only one sublattice whereas the \{100\} interface studied above
contains the same amount of spin up and down sites.\\
\begin{figure}
\centerline{\epsfxsize=\linewidth\epsfbox{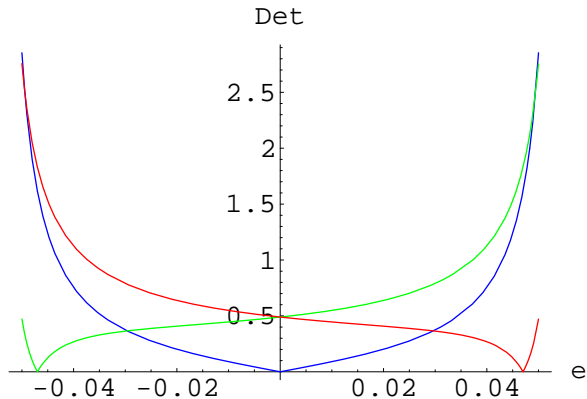}}
\caption{Determinant of $\alpha_r$ versus the energy $\epsilon$ for
  the \{110\} AF/dSC interface. Again this is plottet inside the
  superconducting gap and with $q=0.1$. The dashed curve is the usual
  case of an I/dSC 
  interface which clearly contains a ZEBS (the insulator state is
  obtained by performing the substitution $M_n \rightarrow -M_n$ for the hole
  part of the BdG equations only). The solid curves show the
  spin splitting of the ZEBS for this particular value of $q$.}   
\end{figure}
The splitting of the ZEBS by $\Delta^s$-mixing in the usual situation of a
I/dSC interface has been extensively studied in the
literature\cite{fogelstrom,sigrist,honerkamp}. It is also well-known that a magnetic
field further splits the ZEBS\cite{covington}. The above spin
splitting at AF/dSC interfaces is similar to this magnetic field
effect in the sense that the magnetic interface effectively acts as a local
magnetic field. A similar effect caused by a correlation induced
magnetization near the interface in the case of a I/dSC surface was
discussed by Honerkamp {\sl et.al}\cite{honerkamp}. This ``Zeeman'' effect is
also directly related to the split zero energy Andreev mode observed in the
center of vortex cores of underdoped cuprates where local antiferromagnetism
has been shown to exist\cite{pan,renner,andersen,zhu,miller,halperin}.\\
To the best of our knowledge there has been
no self-consistent calculation investigating any \{110\} AF/dSC interface induced
subdominant order parameters. 
\begin{figure}
\centerline{\epsfxsize=\linewidth\epsfbox{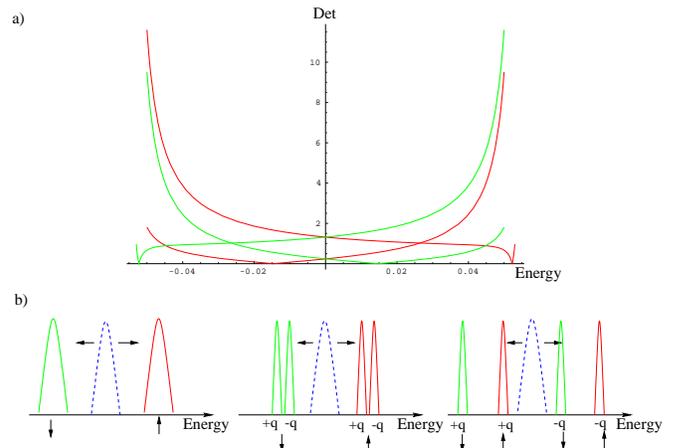}}
\caption{a) Same as in Figure 2, but with an induced extended s-wave
  gap function near the interface, i.e. $d \rightarrow d+is$. For
  clarity we do not show the original ZEBS (dashed curve from Fig. 2). b)
  Schematic representation of the splitting of the original zero
  energy Andreev bound state (dashed curve): 
  1) The antiferromagnetic interface breaks the spin
  degeneracy, as shown in Fig. 2. 2) Induction of a possible
  sub-dominant s-wave gap 
  parameter $\Delta^s$ further splits the spin up/down states by breaking the
  directional degeneracy. 3) Only when $\Delta^s$ exceeds a critical
  value is an interface current induced. In this last figure, which
  corresponds to the situation from a),
  $\Delta^s$ is equal to $\Delta^d$ on the interface and decreases
  linearly to zero within 20 sites of the interface.}   
\end{figure} 
However, we know from the study of I/dSC
surfaces\cite{fogelstrom,matsumoto} that the strong pair breaking
effects of a \{110\} geometry, as opposed to a \{100\} surface,
tends to stabilize the subdominant s-wave component. Thus, even though there is
no Fermi surface instability begging for removal of the ZEBS from the
Fermi level in the case of a AF/dSC \{110\} interface, one should
still consider the effect of an additional local superconducting 
order parameter $is$ competing with the
splitting caused by the magnetism. The consequences of this
competition for the ZEBS are discussed in Figure 3.\\ 
The induction of a surface current is a well-known
consequence of the time reversal symmetry broken state of I/dSC
interfaces\cite{fogelstrom,sigrist}. However, for the AF/dSC interface
with a locally induced $d \pm is$ order parameter   
there is a critical value of $\Delta^s_c$ before a current runs along the
interface\cite{comment2}. In Figure 3a we show the situation when the induced
$\Delta^s$ has exceeded this critical value. Figure 3b is a schematic
representation of the splitting of the original ZEBS with the first
sketch corresponding to the parameters from Fig. 2 and the last sketch
to those from Fig 3a. We stress that only a
self-consistent model calculation can determine the magnitude of
the directional splitting caused by $is$ compared to the spin
splitting caused by the antiferromagnetism, and hence the relevancy
of the interface current.\\
In conclusion we have set up a simple method so determine the
existence of bound states at the interfaces of d-wave superconductors
and antiferromagnets. In particular we studied the
energetics of the notorious zero energy mode bound to \{110\} I/dSC interfaces
first discovered by Hu\cite{hu}. This state is always spin split when
the insulator is an antiferromagnet and is analogous to the split
states found around the magnetic vortex cores of YBCO and BSCCO
crystals. In the case of an array of junctions corresponding to a
periodic domain of vertical or diagonal stripes these states will
hybridize and eventually form a band. A current along the interface exists 
only when the effect of a competing, interface induced $is$ component 
exceeds the spin splitting.

\end{multicols}

\begin{references}
\bibitem{comment} The dGSJ states are usually thought of as existing
  in a narrow normal region (within the superconducting coherence
  length $\xi$) from the surface. However, is is known that these states
  survive in the limit $\xi \rightarrow 0$. See also \cite{lofwander}.
\bibitem{comment2} Here we assume the large effect is the spin
  splitting since $M$ gives rise to the Hubbard gap in the bulk
  antiferromagnet, and $\Delta^s$ is an extra perturbation. In the
  opposite case it would take a critical magnetization $M_c$ to
  prevent the interface current induced by $\Delta^s$.
\end{references}
\end{document}